\documentclass[fleqn,10pt]{wlscirep}

\usepackage{standalone}
\usepackage{amssymb,amsmath,amsthm}%,physics}

\usepackage[utf8]{inputenc}
\usepackage[T1]{fontenc}

\usepackage{graphicx}
\usepackage{adjustbox,subcaption}
\usepackage{xcolor}
\usepackage{tikz}
\usetikzlibrary{backgrounds,calc,decorations.pathmorphing,fadings,scopes,patterns}
\tikzset{>=stealth}
\pgfdeclarelayer{background layer}
\pgfdeclarelayer{foreground layer}
\pgfsetlayers{background layer,main,foreground layer}
\usepackage{color}
\usepackage{framed,mdframed}
\graphicspath{{./fig/}{./}}

\newcommand*{\s}[1]{\ensuremath{_\mathrm{#1}}}
\newcommand*{\up}[1]{\ensuremath{^\text{#1}}}
\newcommand*{\dd}[1]{\ensuremath{\mathrm{d}#1\:}}
\newcommand*{\da}{\dagger}
\newcommand*{\ket}[1]{\ensuremath{\left|#1\right\rangle}}
\newcommand*{\pdv}[2]{\frac{\partial #1}{\partial #2}}
\newcommand*{\ev}[2]{\ensuremath{\langle #2 | #1 | #2 \rangle}}
\newcommand*{\cT}{\mathcal{T}}
\newcommand*{\nocooling}{\ \stackrel{n_0 = n\s{th}}{ = } \ }

\usepackage[capitalize,poorman]{cleveref}
\title{Photon-phonon-photon transfer in~optomechanics}
\author[1,*]{Andrey A. Rakhubovsky}
\author[1]{Radim Filip}
\affil[1]{Department of Optics, Palack{\'y} University, 17. Listopadu 12, 771 46 Olomouc, Czech Republic }
\affil[*]{andrey.rakhubovsky@gmail.com}

\begin{abstract}
	We consider transfer of a highly nonclassical quantum state through an optomechanical system.  That is we investigate a protocol consisting of sequential upload, storage and reading out of the quantum state from a mechanical mode of an optomechanical system.  We show that provided the input state is in a test-bed single-photon Fock state, the Wigner function of the recovered state can have negative values at the origin, which is a manifest of nonclassicality of the quantum state of the macroscopic mechanical mode and the overall transfer protocol itself.  Moreover, we prove that the recovered state is quantum non-Gaussian for wide range of setup parameters. We verify that current electromechanical and optomechanical experiments can test this complete transfer of single photon.
\end{abstract}

\date{\today}
\maketitle

\begin{document}

%  >>>

\section*{Introduction} % <<<
\label{sec:introduction}

Quantum optomechanics~\cite{aspelmeyer_cavity_2014,khalili_quantum_2016} can reversibly interconnect radiation to mechanical oscillators at the level of individual quanta~\cite{thompson_strong_2008,groblacher_observation_2009,verhagen_quantum-coherent_2012,cohen_phonon_2015}
This connection has both fundamental and also practical consequences.
At fundamental side, it controllably converts a single photon to single phonon and then, back to individual photons.
It proves fundamental aspect that the information carried by the physical state of single quantum can be transferred between totally different physical systems.
Also, it says message to quantum thermodynamics~\cite{vinjanampathy_quantum_2016}, that minimal portion of  energy can substantially change its form.
At practical side, such the high-fidelity photon-phonon-photon transducers can be used to sense mechanical motion at levels of single phonon level.
It can stimulate quantum metrology of mechanical motion~\cite{schnabel_squeezed_2016} beyond the current state of the art.
Gaussian quantum entanglement~\cite{palomaki_entangling_2013} and squeezed states~\cite{pirkkalainen_squeezing_2015} between microwave radiation and mechanical oscillator have been experimentally generated and read out using resolved sideband regime.
However, Gaussian states are not sufficient resources for advanced quantum technology.
It requires non-Gaussian quantum states~\cite{mari_positive_2012} to open full space of the applications.
Single photon state is basic non-Gaussian resource for optical technology, therefore the single phonon states are as well basic for non-Gaussian quantum optomechanics.
A mutual interconversion between a photon and a phonon is thus an important proof of this fundamental correspondence.
A single photon is simultaneously a very good test-bed to verify quantum nature of the physical interface between light and mechanics.
Recently, nonclassical correlations of phonon with photon at optical frequencies have been unconditionally generated~\cite{riedinger_non-classical_2016}.  This opens future possibilities to exploit optomechanics operated at single phonon level.
The transducers between optical, mechanical and microwave modes have been considered~\cite{safavi-naeini_proposal_2011, wang_using_2012, tian_adiabatic_2012, tian_optoelectromechanical_2015, zhang_transitionless_2016} and realized in continuous-wave regime~\cite{hill_coherent_2012, andrews_bidirectional_2014, andrews_quantum-enabled_2015,  lecocq_mechanically_2016}.
However, the continuous wave regime with steady states is not suitable when quantum processing and metrology runs in time slots.
In this case, all operations are pulsed with optimized temporal shapes.
Theory proposals of photon-to-phonon transfer in the pulsed~\cite{khalili_preparing_2010, mcgee_mechanical_2013, rakhubovsky_squeezer-based_2016, bennett_quantum_2016} and stroboscopic regime~\cite{hoff_measurement-induced_2016} have been recently analyzed.
Also squeezed light improved read-out of mechanical state has been separately investigated~\cite{filip_transfer_2015}.

In this paper, we analyze complete photon-phonon-photon coherent transfer for microwave and optical experiments in the pulsed and red side-band resolved regime.
We investigate in detail possibility to transfer the negativity of Wigner function and more general quantum non-Gaussianity through an optomechanical system in contrast to studies of engineering of a nonclassical state of a mechanical mode~\cite{rogers_hybrid_2014, teklu_nonlinearity_2015}.
Transfer of both negativity and non-Gaussianity is much more demanding than the transfer of Gaussian entanglement.
We verify that the electromechanical platforms~\cite{palomaki_entangling_2013,ockeloen-korppi_low-noise_2016} with microwave radiation are already feasible for such transduction preserving negativity of Wigner function.
The optomechanical experiments require coupling higher than critical to reach this possibility.
To preserve quantum non-Gaussianity of single photon, optical experiment~\cite{riedinger_non-classical_2016} is however already sufficient.
Our analysis can be extended to other systems~\cite{xia_opto-magneto-mechanical_2014}.
We numerically verify all the predictions using analysis beyond adiabatic elimination of intracavity field.
We use single photon as a basic test bed for photon-phonon-photon channel, however, it verifies basic quality of coherent transfer for other non-Gaussian states.
Our results therefore open path to use photon-phonon-photon transducer for many future experiments in quantum metrology, quantum nonlinear optomechanics and quantum thermodynamics.

% >>>
\section*{Summary of the main results} % <<<
\label{sec:what_happens_in_adiabatic_regime}

% Scheme figure %<<<
\begin{figure*}[htb!]
	\centering
	\includegraphics[width = \linewidth]{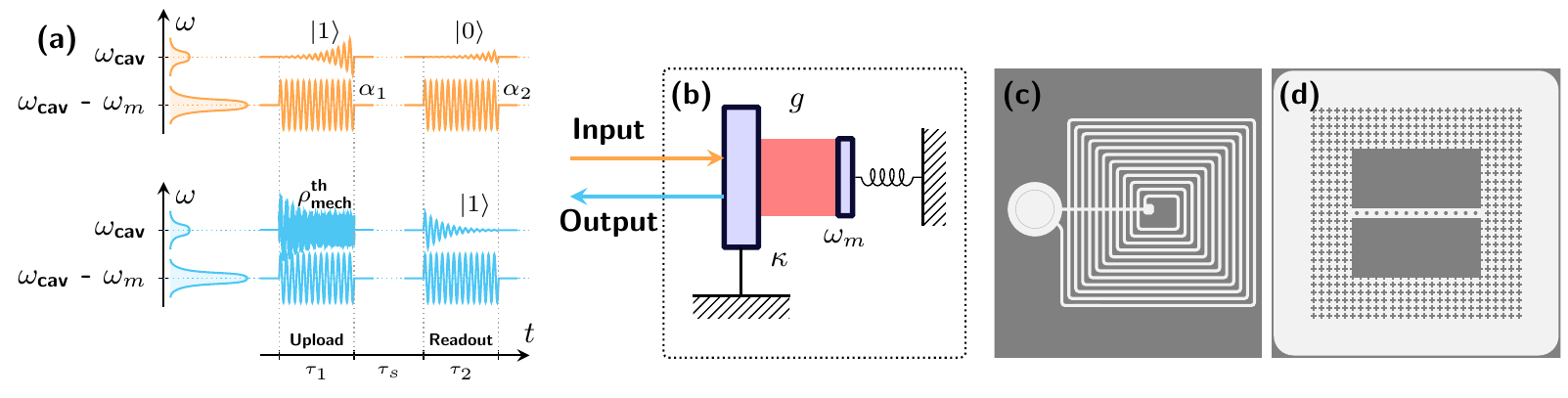}
	{\phantomsubcaption\label{fig:scheme-a}}
	{\phantomsubcaption\label{fig:scheme-b}}
	{\phantomsubcaption\label{fig:scheme-c}}
	{\phantomsubcaption\label{fig:scheme-d}}
	\caption{
		(\subref{fig:scheme-a})~A coherent pulse with amplitude $\alpha_1$ in the resonantly red-detuned pulse enhances the optomechanical coupling and allows to swap the thermal state of mechanical mode $\rho\up{th}\s{mech}$ with the single-photon state $\ket 1$ of the resonant pulse converting the latter to a phonon in the optomechanical system (\subref{fig:scheme-b}).  A consequent readout pulse with amplitude $\alpha_2$ performs a reverse swap to convert the evolved phonon to a photon $\ket 1$ in a mode with exponentially decaying envelope.  Examples of experimental systems include (\subref{fig:scheme-c}), electromechanical system~\cite{palomaki_entangling_2013}, (\subref{fig:scheme-d}), optomechanical crystal~\cite{chan_laser_2011}.
	}
	\label{fig:fig0}
\end{figure*}%>>>

It has been shown previously~\cite{hofer_quantum_2011} that in presence of a strong red-detuned coherent pulse, the quantum states of the mechanical mode of an optomechanical cavity and the resonant with the cavity pulse can in principle be swapped. That is, the quantum state of an incident pulse is transferred to the mechanical mode and the state of the latter is mapped to the reflected light.  The scheme analyzed here uses this state swap twice: first time to upload a highly nonclassical (single-photon) quantum state to the mechanical mode and second time to read the evolved state out (see~\cref{fig:fig0}).  We were inspired by the quality of state read out in experiment~\cite{palomaki_entangling_2013}.

First, the pulse in single-photon state in the optical mode described by annihilation operator $A\up{in}$ is shone upon the cavity at its resonance frequency $\omega\s{cav}$.  A strong classical coherent pulse with amplitude $\alpha_1$ at  frequency $\omega\s{cav} - \omega_m$ enables the state swap, so that the mechanical state after the pulse reads~\cite{hofer_quantum_2011,filip_transfer_2015}
\begin{equation}
	\label{eq:mech_after_upload}
	a_m (\tau) = \sqrt{ T_1 } A\up{in} + \sqrt{ 1 - T_1 } a_m (0),
\end{equation}
where $T_1 = 1 - \exp( - 2 g_0^2 N_1 \tau_1 / \kappa )$ is the transmittivity associated with the partial state swap.  Here $\kappa$ is the cavity decay rate and $g_0$ is the single-photon optomechanical coupling strength enhanced by the mean intracavity photon number $N_1 = | \alpha_1 |^2 $.  If the interaction time $\tau_1$ or optomechanical coupling $g_0 \sqrt{ N_1 }$ is sufficient, the optical state is perfectly mapped to the mechanical mode.  In practice, high optomechanical coupling may be inaccessible, or cause unwanted heating due to absorption in mirrors.  The temporal duration~$\tau_1$ of the pulse is as well limited by mechanical decoherence.

After the upload, the mechanical state is left to evolve for time $\tau\s{s}$.  During this time the mechanical environment admixes thermal noise to the quantum state of the mechanical mode.  This effect can be described by an effective transmittance $\delta = e^{ - \gamma \tau\s{s}}$, where $\gamma$ is the mechanical damping rate.  This admixture happens as well during the upload stage (and consequent readout), but we omit it for the state swap stages as this effect can be reduced by shortening the pulses durations.

Finally, another state swap is performed to read the mechanical state out.  In a full analogy with~\cref{eq:mech_after_upload} the state of the output optical mode reads
\begin{equation}
	A\up{out} = \sqrt{ T_2 } a_m (\tau_1 + \tau\s{s}) - \sqrt{ 1 - T_2 } A\up{vac},
\end{equation}
where $A\up{vac}$ describes a vacuum mode incident to the cavity, $T_2 = 1 - \exp( - 2 g_0^2 N_2 \tau_2 / \kappa )$, $N_2 = | \alpha_2 |^2$ and $\tau_2$ is the temporal duration of the readout pulse.

The complete channel from the input optical state with annihilation operator $A\up{in}$ to the output one with $A\up{out}$ will be degraded by optical losses caused by e.g. imperfect coupling to the optomechanical cavity.  The optical loss performs admixture of vacuum to the signal mode and is characterized by transmittivity $\eta =  \kappa_e / \kappa $, where $\kappa$ and $\kappa_e$ are respectively total cavity decay rate and the cavity decay rate due to the coupling to the waveguide.

All the steps constituting the channel thus can be described by known transmittivities and therefore the quadratures of the output optical state can be expressed in terms of quadratures $Q\up{in}$ of input state and quadratures of added noise $Q\up{N}$
\begin{gather}
	\label{eq:an_definition}
	Q\up{out} = \sqrt{ \cT } Q\up{in} + \sqrt{ 1 - \cT } Q\up{N},
	\quad
	Q = X, Y,
	\\
	X^k = A^k{}^\da + A^k,
	\quad
	Y^k = i ( A^k{}^\da - A^k ).
\end{gather}
Here the transmittivity equals $\cT = T_1 T_2 \eta^2 \delta$.

The noisy mode $Q\up{N}$ in the equation above comprises optical noises coming from imperfect coupling, thermal mechanical noise and the initial mechanical state.  We assume all these modes to be zero-mean Gaussian noises with known variances.  From the latter we construct the covariance matrix of $Q\up{N}$ which is a diagonal $2\times 2$ matrix with the diagonal elements equal
%equal to $\cm V\s{N} = \openone_{2 \times 2} V\s{N}$ with
\begin{equation}
	\label{eq:adiabatic_noise_variance}
	% V\s{N} =
	% % \\
	%
	V_N
	=
	1
	+ \frac{ T_2 \eta }{ 1 - \cT } \Big[ 2 n_0 \delta ( 1 - T_1) + 2 n\s{th} ( 1 - \delta )  \Big] \nocooling
	1 + 2 n\s{th} \frac{ \eta T_2 ( 1 - T_1 \delta ) }{ 1 - \cT }.
\end{equation}
Here we assumed that the mechanical mode and its environment are initially in thermal states with occupation $n_0$ and $n\s{th}$ respectively and all the auxiliary optical modes are in vacuum.

% Added noise variance figure %<<<
\begin{figure*}[tb!]
	\centering
	\includegraphics[width=\linewidth]{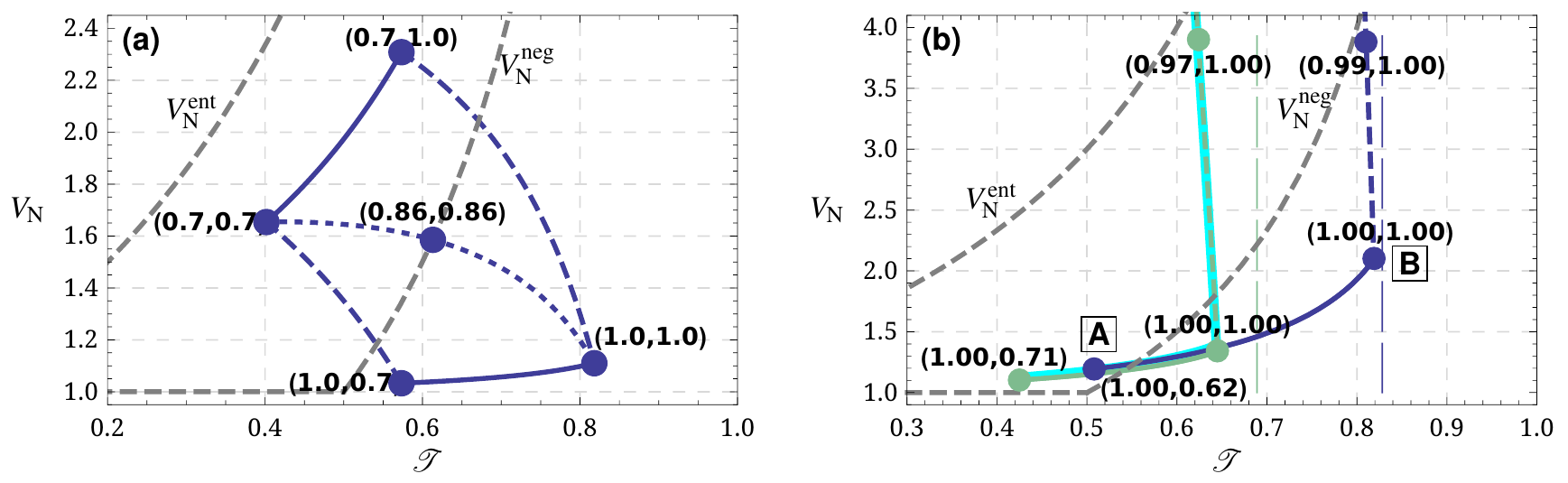}
	{\phantomsubcaption\label{fig:voft-a}}
	{\phantomsubcaption\label{fig:voft-b}}
	\caption{Added noise variance $V\s{N}$ (in shot noise units) as a function of the optomechanical channel transmittivity $\cT$ for (a): the adiabatic regime (\cref{eq:adiabatic_noise_variance}) and for (b) full solution (see Methods).  Along solid [dashed] lines $T_1$ [$T_2$] is constant, along the dotted line $T_1 = T_2$.  Values in brackets show pairs $(T_1, T_2)$ at the corresponding points.  Gray dashed lines show threshold variances that allow respectively transfer of negativity of a Wigner function ($V\up{neg}\s{N} = \cT / ( 1 - \cT )$) and preservation of entanglement ($V\up{ent}\s{N} = 2 \cT / ( 1- \cT ) + 1 $)~\cite{serafini_quantifying_2005}.  Vertical dashed lines in (b) show the maximum possible transmittance allowed by coupling efficiency $\cT = \eta^2$.
		To justify validity of the Rotating Wave Approximation (RWA) used throughout the paper, we plot with solid cyan lines the solution obtained without using the RWA for one of the sets of parameters.
	}
	\label{fig:voft}
\end{figure*}
% >>>

In order to transfer negativity of the Wigner function of a single-photon state, the variance \cref{eq:adiabatic_noise_variance} should satisfy~$V\s{N} < V\s{N}\up{neg} \equiv \cT / ( 1 - \cT )$.  Our goal is thus to increase the transmittivity $\cT$ preserving moderate added noise variance $V\s{N}$.  From the expression for $\cT$ it is evident that it can be increased by increasing either of $T_1$ and $T_2$.  However, increasing $T_1$ not only increases $\cT$ but decreases $V\s{N}$ by diminishing the ratio of the initial thermal mechanical state in the output.  Increase of $T_2$ is not so crucially important and it causes increase of both $\cT$ and $V\s{N}$.  These simple considerations can be proven by calculating partial derivatives
\begin{align}
	\left( \pdv{ V\s{N} }{ \cT } \right)_{T_2}
	& =
	- 2 n_0 \frac{ 1  - \delta  \eta ^2 T_2}{\eta  (1 - \cT)^2}
	+ 2 n\s{th} \frac{ (1 - \delta) \eta  T_2}{(1 - \cT)^2}
	\nocooling -  2 n\s{th} \frac{ 1 - \eta^2 T_2 }{ \eta ( 1 - \cT )^2 } \leqslant 0,
	\\
	\left( \pdv{ V\s{N} }{ \cT } \right)_{T_1}
	& =
	2 n_0 \frac{ 1 - T_1 }{\eta  ( 1 - \cT)^2 T_1} + 2 n\s{th} \frac{ 1 - \delta }{\delta  \eta  ( 1 - \cT)^2 T_1}
	\nocooling 2 n\s{th} \frac{ 1 - T_1 \delta }{ T_1 \eta \delta ( 1 - \cT )^2 } \geqslant 0.
\end{align}

% The nongaussianity Fig %<<<
\begin{figure*}[tb!]
	\centering
	\includegraphics[width=\linewidth]{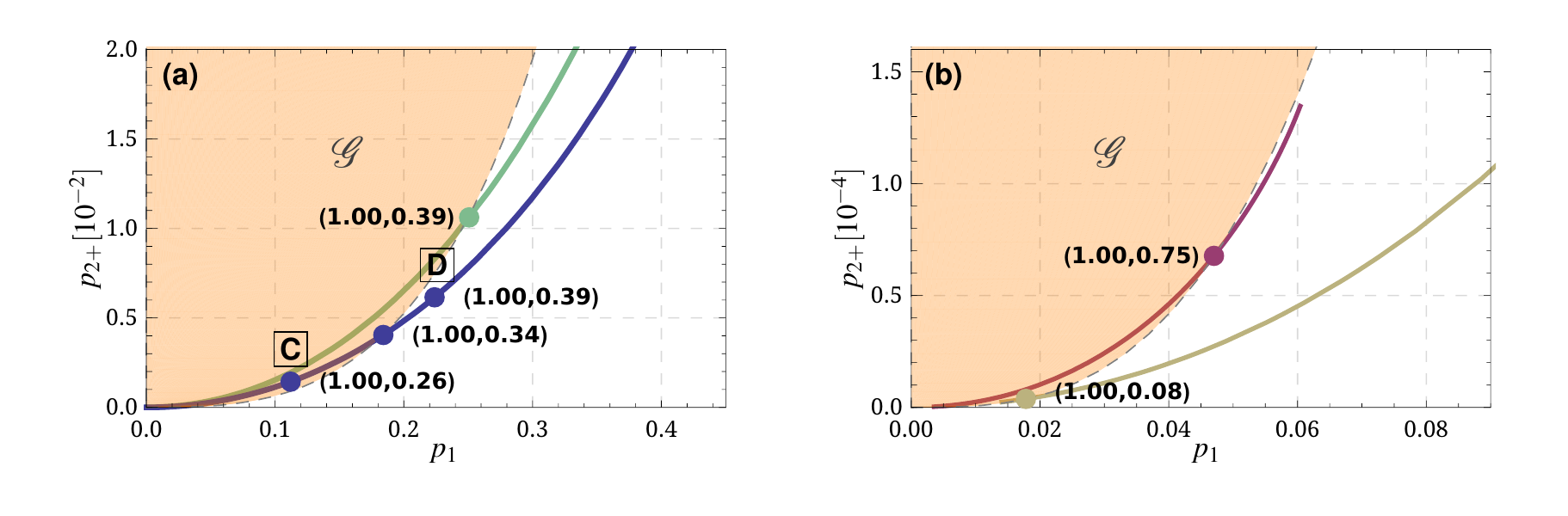}
	% \begin{adjustbox}{max totalsize={1\linewidth}{1\textheight},center}
	% 	\input{fig/fig3.tex}
	% \end{adjustbox}
	{\phantomsubcaption\label{fig:ng-a}}
	{\phantomsubcaption\label{fig:ng-b}}
	\caption{
		The probability of a multiphoton contribution $p_{2+} \equiv 1 - p_0 - p_1$ versus the probability of a single-photon state $p_1$ in the read out state of electromechanical systems~(\subref{fig:ng-a}) and optomechanical system~(\subref{fig:ng-b}).  The plots in~(\subref{fig:ng-a}) are made for parameters of electromechanical experiments~\cite{palomaki_entangling_2013} (green lines) and~\cite{ockeloen-korppi_low-noise_2016} (blue lines).  The plots in~(\subref{fig:ng-b}) are made for the critically coupled $\eta = 0.5$ (yellow line) and undercoupled $\eta = 0.25$ (purple line) optomechanical cavity~\cite{riedinger_non-classical_2016}.  In brackets are pairs of transmittances $(T_1,T_2)$ that mark the corresponding line entering non-Gaussianity area.
	}
	\label{fig:ng_prob}
\end{figure*}%>>>

The principal dependence of $V\s{N}$ on $\cT$ is shown for different transmittances $T_{1,2}$ in~\cref{fig:voft}~(a).  The mechanical mode is considered to be at equilibrium with the environment at $n_0 = n\s{th} = 1$.  Even this low occupation creates a demand for a very efficient state upload, so that increasing $T_1$ with constant $T_2$ helps to suppress the added noise dramatically (dashed lines).  It is worth noting that for certain values of $T_1$ the successful transfer of the negativity is impossible regardless of $T_2$.  Note, it is evident that is much easier to reach entanglement preserving channel than to transfer negative Wigner function.  From the~\cref{fig:voft}~(b) that shows the analysis following the full solution (see~Methods) we conclude that the current electromechanical setups~\cite{palomaki_entangling_2013,ockeloen-korppi_low-noise_2016} are capable of the transfer of the negativity of the Wigner function provided there is a single-photon Fock state at input.  The correspondence between~\cref{fig:voft}~(a) and~\cref{fig:voft}~(b) allows us to conclude that the simple model based on the effective transmittances provides qualitatively correct understanding of the physics of the transfer.

A major impediment to the successful photon-phonon-photon transfer preserving the negativity of Wigner function is the initial occupation of the mechanical mode.  The influence of the mechanical bath during the state swap stages can in principle be suppressed by taking shorter pulses, however, the nonzero initial occupation has to be compensated for either by precooling of the mechanical mode or by a very efficient upload.  In spite of our considered experimental scheme the precooling happens to be an additional resource that can relax the requirements to the upload of the quantum state stage.  The considerations without the additional precooling therefore happen to be more conservative as such regime is more challenging experimentally and it requires the  upload efficiencies close to unity.  These efficiencies, however, are realistic at least in the domain of state-of-the-art electromechanics~\cite{palomaki_entangling_2013}, where the transmittances can reach level as high as $T_1 > 1 - 10^{-5}$.  There is therefore an open way to experimentally test photon-phonon-photon channel preserving negativity of Wigner function.

We should note that in order to transmit negativity of the Wigner function, one should have $\cT > 1/2$ which places restrictions on each of individual transmittances, particularly on the efficiency of the coupling, $\eta$.  That is, critically coupled systems with $\eta = 0.5$ are incapable of transferring the negativity regardless of the effectiveness of state upload and readout.  For these systems we consider another measure of nonclassicality, namely the quantum non-Gaussianity.  A convenient criterion~\cite{filip_detecting_2011,jezek_experimental_2012} allows to determine whether a state can not be represented as a mixture of Gaussian states.  The criterion establishes an upper boundary $p_{1,G} (p_0)$ for the probability of the single photon state $p_1$ for a fixed probability of vacuum~$p_0$.  That is if for a certain quantum state the condition $p_1(p_0) > p_{1,G}(p_0)$ is fulfilled, then this state is definitely non-Gaussian.  This boundary can be parametrized as follows:
\begin{equation}
	p_0 = \frac{ e^{ - e^r \sinh r }}{ \cosh r },
	\quad
	p_{1,G} = \frac{ e^{4r } - 1 }{ 4 } \frac{ e^{ - e^r \sinh r }}{ \cosh^3 r },
\end{equation}
with $r \in [0 , +\infty)$.  To evaluate the non-Gaussian properties of the restored state of the field we therefore need to evaluate the probabilities $p_0$ and $p_1$ corresponding to this state and certify if they belong to the region accessible to Gaussian states.  The latter is denoted as $\mathcal G$ and shaded at~\cref{fig:ng_prob}.  Experimentally both probabilities can be determined from homodyne detection scheme used to verify negativity of Wigner function~\cite{filip_detecting_2011}.  Alternatively, single photon counters can be used~\cite{straka_quantum_2014}.

\begin{figure}[htb]
	\centering
	\includegraphics[width=.9\linewidth]{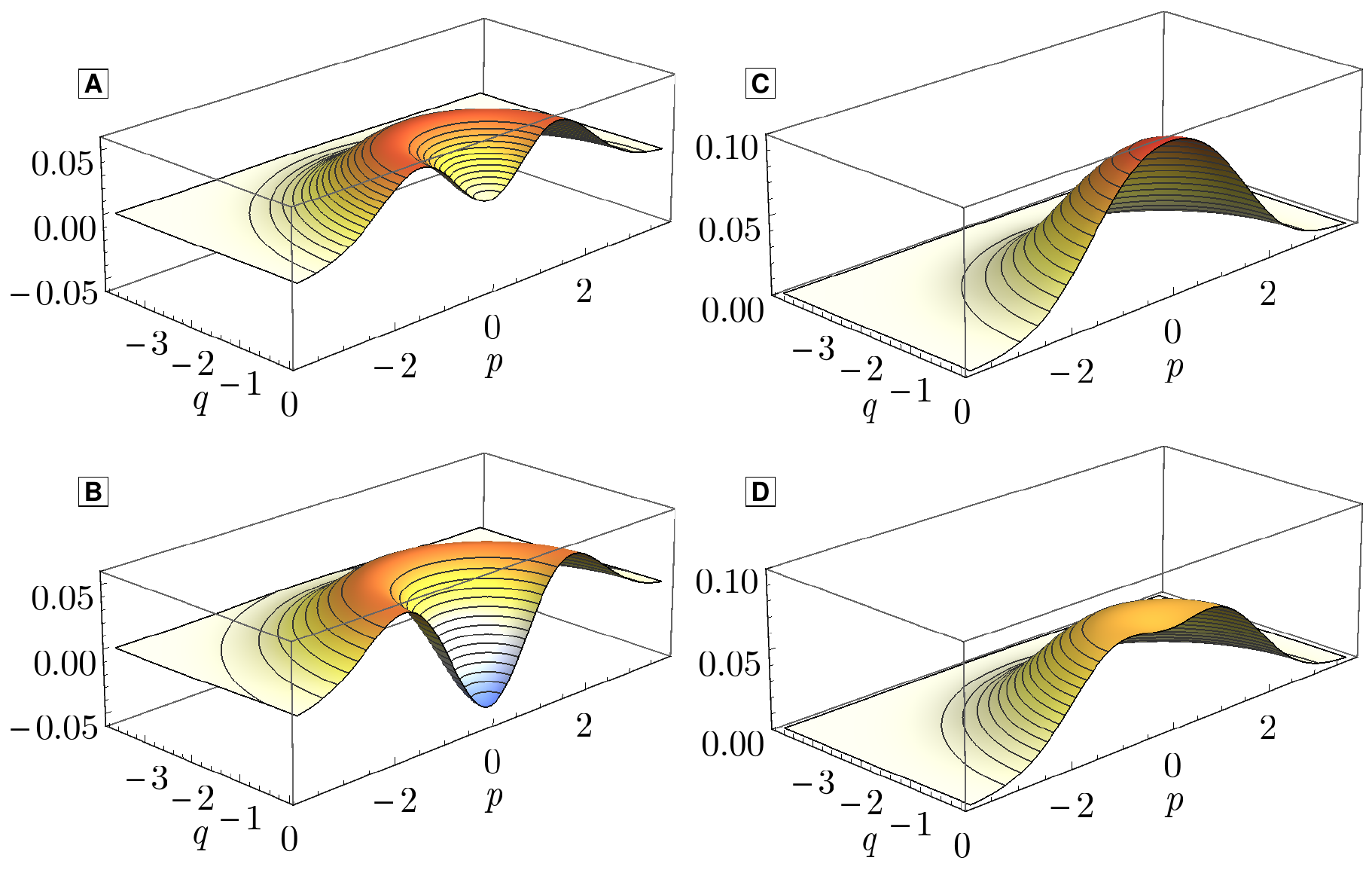}
	\caption{Wigner functions of the restored state for the points before (A) and after (B) crossing the threshold $V\up{neg}\s{N}$ (Fig.~\ref{fig:voft}), and outside (C) and inside (D) the non-Gaussianity area $\mathcal G$ (Fig.~\ref{fig:ng_prob}).  All plots are made for the parameters of the experiment reported in~\cite{ockeloen-korppi_low-noise_2016}.
}
	\label{fig:fig4}
\end{figure}

In order to evaluate $p_0$ and $p_1$, we use our estimates for the transmittance $\cT$ and added noise $V\s{N}$ to calculate the Wigner function of the recovered state, from which we find the needed probabilities.  Results of this calculation are visualized for recent electro- and optomechanical experiments at~\cref{fig:ng_prob}.  Example Wigner functions are demonstrated in Fig.~\ref{fig:fig4}.  The demonstration of quantum non-Gaussianity transferred by a photon-phonon-photon channel is feasible with both platforms.  The required efficiencies, particularly ones of the readout, are significantly lower for the electromechanics than that needed for the transfer of negativity of Wigner function.  Note that for the weakly coupled optomechanical system, the transfer of negativity is impossible regardless of the efficiencies of the state swap.  Due to the initial occupation of the mechanical modes a very efficient state upload is required.

% >>>
\section*{Methods} % <<<
\label{sec:what_happens_in_full_solution}

For a rigorous analysis that accounts for the imperfections properly, we start with the linearised Heisenberg-Langevin equations for the optomechanical system with a strong resonantly red-detuned pump.  In terms of annihilation operators $a_c$ and $a_m$ of optical and mechanical modes respectively the equations read~\cite{filip_transfer_2015}:
\begin{align}
	\label{eq:pulse_HL:1}
	\dot a_c & = - \kappa a_c - g a_m + g a_m^\da e^{ 2 i \omega_m t } + \sqrt{ 2 \kappa }
	\left( \sqrt{ \eta } a\up{in} (t) + \sqrt{ 1 - \eta} a\up{vac} (t)\right),
	\\
	\label{eq:pulse_HL:2}
	\dot a_m & = - \frac \gamma 2 a_m + g a_c + g a_c^\da e^{ 2 i \omega_m t } + \sqrt{ \gamma } a\up{th} (t).
\end{align}
Here $\omega_m$ is the mechanical mode frequency, $\gamma$ is mechanical viscous damping coefficient, $a\up{in}$ and $a\up{vac}$ are the input and auxiliary vacuum optical modes respectively, and $a\up{th}$ describes the Markovian noises from the mechanical environment.  The coupling equals $g = g_k \equiv g_0 \sqrt{ N_k}$ for $k$-th pulse and $g = 0$ during the storage between the pulses.  Usually the condition of resolved sideband $\kappa \ll \omega_m$ holds and therefore the rotating wave approximation (RWA) can be applied thus eliminating the rapid terms proportional to $\exp(2 i \omega_m t )$ in the equations above.  To prove validity of our results obtained with help of RWA, on Fig.~\ref{fig:voft} we present a solution that takes into account the rapid terms by the perturbation method.  The fact that this complete solution coincides with the solution taking advantage of RWA justifies use of the latter for the rest of the paper.

With help of the input-output relation
\begin{equation}
	a_2\up{out} (t) = - a_2\up{in} (t) + \sqrt{ 2 \kappa_e } a_c (t),
\end{equation}
we are able to construct the relation connecting the bosonic operator of the input optical mode $A\up{in}$ with the output $A\up{out}$ in analogy with~\cref{eq:an_definition}.

The bosonic operators are defined as
\begin{equation}
	A\up{k} = \dfrac{ \int_0^\tau a\up{k} (t) f\up{k} (t) dt }{ \sqrt{ \int_0^\tau | f\up{k} (t) | ^2 dt }},
	\quad
	\text{ k = in,out }.
\end{equation}
The pulse shapes $f\up{in}(t)$ and $f\up{out}(t)$ are determined from the~\cref{eq:pulse_HL:1,eq:pulse_HL:2}.  Generally expressions for these functions are quite complicated hence we do not write those explicitly here.  In the limit of high $\kappa \gg g,\gamma$, where the cavity mode can be approximately eliminated~\cite{hofer_quantum_2011},
\begin{equation}
	A\up{in} = \sqrt{\frac{ 2 G_1 }{ e^{ 2 G_1 \tau_1 } - 1 }} \int_{\tau_1} \dd{t} a_1\up{in} (t) e^{ G_1 t },
	\quad
	A\up{out} = \sqrt{\frac{ 2 G_2 }{ 1 - e^{ - 2 G_2 \tau_2 }}} \int_{\tau_2} \dd{t} a_2\up{out} (t) e^{ - G_2 t },
\end{equation}
with $G_i = g_0^2 N_i / \kappa$.

Knowing the statistics of each individual constituent of the noisy mode $Q\up{N}$, we can evaluate the variance of the added noise: $\text{var}\:{ X\up{N}} = \text{var}\:{Y\up{N}} = V\s{N}$ e.g., following the steps outlined in~\cite{vostrosablin_pulsed_2016}, Appendix B.  It is visualized as a function of the transmittivity $V\s{N} (\cT)$ in \cref{fig:voft}~(b).

The negativity- and entanglement-preserving thresholds $V\up{neg}\s{N}$ and $V\up{ent}\s{N}$ can be easily derived as follows.  The value of $V\up{neg}\s{N}$ is naturally obtained from the value of the Wigner function in the origin of the single-photon that passed the beamsplitter.  The $V\up{ent}\s{N}$ is derived from taking the maximum over the two-mode squeezed vacuum in a thermal channel.  These boundaries are derived carefully in~\cite{serafini_quantifying_2005}.

The effective transmittivities $T_1$ and $T_2$ associated with respectively upload and readout stages present a convenient way to parametrize this dependence.  Each of these transmittivities can be increased via increase of either corresponding pump power (intracavity photon number) or pulse duration.  The latter way is accompanied by increase of the mechanical environment impact.  In the~\cref{fig:voft}~(b) we assume the coupling close to the highest reported and change $\tau_{1,2}$ to produce change of $T_{1,2}$.  We use the following dimensionless parameters to produce the plots:  for the green lines (as reported in~\cite{palomaki_entangling_2013}): $\tau_1 = \tau_2 \leq 100 / \kappa$, $\omega_m \approx 50 \kappa$, $g_1 = g_2 = 0.25 \kappa$, $\tau_s = 3 / \kappa$, $\gamma = 2 \times 10^{-4} \kappa$, $n\s{th} = n_0 = 20$, $\eta = 0.83$.  For the blue lines (following~\cite{ockeloen-korppi_low-noise_2016}) $\tau_1 = \tau_2 \leq 10^3/ \kappa$, $g_1 = g_2 = 0.07 \kappa$, $\tau_s = 100 / \kappa$, $\gamma = 2 \times 10^{-5} \kappa$, $n\s{th} = n_0 = 18$, $\eta = 0.91$.

Evaluation of the transfer of quantum nongaussianity requires computation of the probability of vacuum and of single photon.  If the restored optical quantum state is $\rho_f$, these probabilities equal $p_0 = \ev{\rho_f}{0}$ and $p_1 = \ev{\rho_f}{1}$ and can be determined by homodyne tomography~\cite{filip_detecting_2011}.  The quantum state $\rho_f$ is found as a result of a single photon $\ket 1$ passing a virtual beamsplitter with transmittance $\cT$ that adds noise with variance $V\s{N}$.  The curves in~\cref{fig:ng-a} are produced for same numerical parameters as~\cref{fig:voft}~(b).  To produce~\cref{fig:ng-b} we use parameters reported in~\cite{riedinger_non-classical_2016}, $\tau_1 \leq 3 \times 10^3 / \kappa$, $\tau_2 \leq 2 \times 10^3 / \kappa$, $g_1 = g_2 = 0.04 \kappa$, $\tau_s = 100 / \kappa$, $\gamma = 7 \times 10^{-6} \kappa$, $n\s{th} = n_0 = 1$, $\eta = 0.5$ and $\eta = 0.25$ respectively for yellow and purple lines.

% >>>
\section*{Acknowledgements} % <<<
\label{sec:acknowledgements}

The authors acknowledge financial support from Grant No. GB14-36681G of the Czech Science Foundation.  A.R acknowledges institutional support of the Faculty of Science, Palack{\'y} University.
% >>>
\section*{Author contributions} % <<<
\label{sec:author_contributions}

Both A.R. and R.F. contributed jointly to theory.  R.F. proposed the theoretical concept.  A.R. carried out the detailed calculations. Both authors contributed to the interpretation of the work and the writing of the manuscript.
% >>>
\section*{Additional Information} % <<<
\label{sec:additional_information}

The authors declare no competing financial interests.
% >>>
\bibliography{transfer}

\end{document}